\documentclass[sigconf]{acmart}
\usepackage{CJK}
\usepackage{inputenc}
\usepackage{fontenc}
\usepackage{textcomp}
\usepackage{xcolor}
\usepackage{colortbl}
\usepackage{multirow}
\usepackage{soul}
\usepackage{hyperref} 
\usepackage{subfig}
\usepackage{stfloats}

\AtBeginDocument{%
  \providecommand\BibTeX{{%
    \normalfont B\kern-0.5em{\scshape i\kern-0.25em b}\kern-0.8em\TeX}}}

\copyrightyear{2024} 
\acmYear{2024} 
\setcopyright{acmlicensed}\acmConference[CHI '24]{Proceedings of the CHI Conference on Human Factors in Computing Systems}{May 11--16, 2024}{Honolulu, HI, USA}
\acmBooktitle{Proceedings of the CHI Conference on Human Factors in Computing Systems (CHI '24), May 11--16, 2024, Honolulu, HI, USA}
\acmDOI{10.1145/3613904.3642194}
\acmISBN{979-8-4007-0330-0/24/05}

\begin{document}
\begin{CJK}{UTF8}{gkai}

\title[Discursive Strategies of Gender Debate in Feminist Advocacy]{Persuasion or Insulting? Unpacking Discursive Strategies of Gender Debate in Everyday Feminism in China}

\author{Yue Deng}
\affiliation{%
  \institution{Department of Computer Science and Engineering, Hong Kong University of Science and Technology}
  \city{Hong Kong SAR}
  \country{China}
}
\email{ydengbi@cse.ust.hk}

\author{Zheng Chen}
\affiliation{%
  \institution{Department of Statistics and Actuarial Science, University of Hong Kong}
  \city{Hong Kong SAR}
  \country{China}
}
\email{u3577077@connect.hku.hk}

\author{Changyang He}
\affiliation{%
  \institution{Department of Computer Science and Engineering, Hong Kong University of Science and Technology}
  \city{Hong Kong SAR}
  \country{China}
}
\email{cheai@cse.ust.hk}

\author{Zhicong Lu}
\affiliation{%
  \institution{Department of Computer Science, City University of Hong Kong}
  \city{Hong Kong SAR}
  \country{China}
}
\email{zhicong.lu@cityu.edu.hk}

\author{Bo Li}
\affiliation{%
  \institution{Department of Computer Science and Engineering, Hong Kong University of Science and Technology}
  \city{Hong Kong SAR}
  \country{China}
}
\email{bli@cse.ust.hk}

\renewcommand{\shortauthors}{Deng, et al.}

\newcommand{\gr}{\leavevmode\color{lightgray}}

\begin{abstract}

Speaking out for women's daily needs on social media has become a crucial form of everyday feminism in China. Gender debate naturally intertwines with such feminist advocacy, where users in opposite stances discuss gender-related issues through intense discourse. The complexities of gender debate necessitate a systematic understanding of discursive strategies for achieving effective gender communication that balances civility and constructiveness. To address this problem, we adopted a mixed-methods study to navigate discursive strategies in gender debate, focusing on 38,636 posts and 187,539 comments from two representative cases in China. Through open coding, we identified a comprehensive taxonomy of linguistic strategies in gender debate, capturing five overarching themes including derogation, gender distinction, intensification, mitigation, and cognizance guidance. Further, we applied regression analysis to unveil these strategies' correlations with user participation and response, illustrating the tension between debating tactics and public engagement. We discuss design implications to facilitate feminist advocacy on social media.

\textbf{Content Warning}: This paper contains discussions on gender debate that may include swear words and sensitive topics, such as sex, potentially causing discomfort.

\end{abstract}



\maketitle

\section{INTRODUCTION}

\begin{quote}
    \textit{``Go away you man! What the hell is it to you?'' --- post by a woman}
\end{quote}

\begin{quote}
    \textit{``All these women are acting like big babies and still have the nerve to act self-righteous!'' --- post by a man}
\end{quote}

Everyday feminism is a branch of feminism that encompasses a wide range of practices aimed at empowering individuals to challenge gender inequalities in their daily lives \cite{schuster2017personal}. It serves as a valuable channel for raising public awareness about feminist issues \cite{kelly2015feminist}. In China, everyday feminism has emerged as a significant approach to feminist advocacy \cite{yin2021intersectional,zhang2023reflections}. However, it faces challenges such as limited visibility and stigmatization under its unique socio-cultural contexts \cite{liu2008cyberactivism,mao2020feminist}, where traditional norms related to male dominance continue to impact public perceptions \cite{han2018searching, mingmo}. The emergence of social media has accelerated the growth of everyday feminism in China, facilitating the sharing of personal experiences and expression of viewpoints, and fostering a vibrant community for feminist information and communication \cite{pruchniewska2019everyday}. It enables ordinary users, who may not necessarily identify as feminists, to not only voice their concerns and advocate for women's daily needs, but also engage in an interactive dialogue (e.g., commenting, and sharing with others) \cite{mao2020feminist}. This trend highlights the potential of social media as a platform for advocating feminist issues in China.
  
While everyday feminism on social media holds great importance, it often intertwines with the rise of \textbf{gender debate}, which manifests as an argument among individuals or groups about gender-related issues \cite{peng2022digital,yutong2023hostility}. Uncivil discourse like offensive, abusive and vicious language, is a significant and obvious component in gender debate \cite{yutong2023hostility,yang2023rural}. It has severe consequences such as polarizing views of issues \cite{anderson2014nasty}, redirecting focus from real gender issues to sheer antagonism \cite{wu2019made}, threatening social harmony and gender equality \cite{yang2022research}, and diminishing citizens' sense of well-being \cite{yutong2023hostility}. Nevertheless, previous research has demonstrated that debate not only has negative impacts but also yields positive outcomes, such as improving collective wisdom \cite{navajas2018aggregated}, enhancing communication skills \cite{bellon2000research}, and promoting reflection on complex social problems \cite{fiesler2019ethical}. Such an effect of a ``double-edged sword'' necessitates a comprehensive understanding of gender debate within everyday feminism to facilitate better communications for gender issues and feminist advocacy on social media. 


When engaging in gender debate, individuals employ diverse language tactics to express viewpoints, support arguments or counter opposing perspectives \cite{cosper2022patterns,aiston2022argumentation}. These debating strategies are referred to as ``\textbf{discursive strategies}'' in our work. Previous works demonstrated that exploring linguistic practices could gain a profound understanding of how individuals engage in online discussions \cite{cavazza2014swearing, sagredos2022slut}, present their arguments \cite{tan2016winning}, and endeavor to influence others' perspectives and reflect identities \cite{rho2018fostering}. Investigating discursive strategies within gender debate could facilitate understanding and communication across genders, nurture the development of constructive debating skills, and advance feminist advocacy \cite{mao2020feminist, cui2022comparative,peng2020feminist}. Furthermore, the effects of discursive strategies also deserve attention. Effective use of discursive strategies empowers women to express their needs, reduces user disengagement caused by conflicts and harmful speech within communities (e.g., leaving the community \cite{sarker2023automated}), and fosters a positive and productive atmosphere \cite{mueller2016positive,roy2013feminist,wang2019chinese}. However, inadequate utilization of strategies may result in polarization and even the ``backfire'' effect (i.e., supporting original opinions
even more strongly \cite{nyhan2010corrections, jiang2018linguistic}). Hence, it is crucial to explore discursive strategies in gender debate within everyday feminism due to the considerable variance in their effects. 

Prior research on feminism in HCI community predominantly focused on integrating feminism theories and perspectives into research and design (i.e., feminist HCI) \cite{bardzell2010feminist, fox2017imagining,tuli2018learning,schlesinger2017intersectional,d2020personal}. Built upon that, researchers also made the efforts to address specific gender-related issues, such as combating sexual abuse \cite{ahmed2014protibadi,andalibi2018social,kannabiran2011hci}, empowering low-income rural women \cite{shroff2010towards,sultana2018design}, and advocating for abortion rights through digital storytelling \cite{michie2018her}. However, gender debate on social media, particularly that addressing everyday needs of women, remains underexplored in HCI literature. While existing works on gendered discourse delved into gendered conflict which usually highlighted one-sided attacks against women \cite{sunden2018shameless,esposito2021dare,herring2002searching}, our work reveals a bidirectional nature of gender debate. Furthermore, given that previous investigation into discursive strategies in gendered discourse was mostly theory-driven (e.g., the Discourse-Historical Approach (DHA) \cite{reisigl2017discourse}), exploring discursive strategies through the lens of user-generated content on social media could offer fresh insights into gender debate. Nevertheless, which discursive strategies are naturally developed by users in gender debate is less explored. Additionally, how these strategies impact user participation and user response is also underinvestigated. To this end, we propose the following research questions:

\begin{itemize}
  
  \item \textbf{RQ1}: What are the user-developed discursive strategies in gender debate on social media?

  \item \textbf{RQ2}: What are the effects of these strategies in gender debate on user participation and user response?

\end{itemize}

To answer these questions, we collected 38,636 posts related to gender debate with 187,539 comments from Weibo, one of China's largest social platforms, and conducted a mixed-methods study to investigate the discursive strategies in gender debate and their effects on user participation and user response. Through text classification in preliminary context exploration, we discovered that gender debate involved a substantial amount of uncivil and constructive discourses, with a noteworthy portion of posts displaying a blend of uncivil and constructive content, highlighting the intricate nature of gender debate. Through an open coding approach (RQ1), we captured users' discursive strategies in gender debate, such as \textit{role reversal} and \textit{gender exclusion} strategies to distinguish genders, as well as \textit{gender-related educating} and \textit{evidence informing} strategies to guide cognizance. By quantifying user response in comments with text classifiers and applying regression analysis (RQ2), we systematically uncovered the correlations between these strategies and user participation (i.e., likes, comments and forwards) as well as different aspects of user response (i.e., incivility, constructiveness, and stance). We found that (i) the strong engagement promotion of gender-oriented strategies though not in a high volume (e.g., \textit{role reversal} strategy), (ii) potential backfiring of seemingly helpful strategies (e.g., \textit{suggestion} strategy), and (iii) particular strategies' divergent correlations with user response in different stances (e.g., \textit{recontextualization} and \textit{overgeneralization} strategies). Based on the findings, we discussed implications for a deeper understanding of gender debate and constructive discussions about gender issues on social media.

This work makes the following contributions to gender debate within everyday feminism in HCI community: (1) we deepened the understanding of gender debate within everyday feminism in China; (2) we collected and labeled a valid dataset of gender debate on social media, serving as a valuable resource for future HCI research in gender-related discussions; (3) we captured a comprehensive taxonomy of user-developed strategies in gender debate; (4) we revealed how different debating strategies correlated with user participation and user response. This work offers valuable insights into discursive strategies that could contribute to the understanding of gendered conversations, while also facilitating a broader exchange of diverse gendered viewpoints on feminist advocacy.

\section{RELATED WORK}
\subsection{Everyday Feminism on Social Media in Chinese Internet Spaces}
Social media is a constant companion for many in everyday feminist practices \cite{pruchniewska2019everyday}. Platforms like Twitter, feminist blogs and Facebook groups extend the reach of feminism into the online world. Women use these platforms to mutually support one another and increase visibility around topics that directly impact their lives \cite{schuster2017personal}. This grassroots approach exemplifies the rise of everyday feminism on social media, driven by collective power and the shared challenging experiences \cite{wu2019made}.


In the United States, there are pivotal events that have played a role in empowering women's voices and addressing their daily needs. One of the most prominent movements is the \#MeToo movement, which gained widespread attention and momentum in 2017 \cite{rho2018fostering, ammari2022moderation, andalibi2018social,gallagher2019reclaiming,hassan2019can}. Through the use of social media platforms, women from diverse backgrounds shared their stories of sexual harassment and assault. However, it also faced criticism for the exclusion of women of color from feminist movements and the overrepresentation of white women \cite{moitra2021parsing,quan2021mapping}. Another significant campaign \#TimesUp emerged as a response to the revelations of sexual misconduct in various industries \cite{mueller2021demographic,choo2019metoo,chawla2021metoo}. This movement emphasized the need to challenge workplace cultures that enable harassment and discrimination. It advocated for reforms in policies, practices, and power dynamics to create safer and more inclusive environments for women. These campaigns sparked important conversations, raised public awareness, and encouraged individuals and institutions to take actions \cite{rho2019hashtag, rodino2018me, castle2020effect}. These feminist practices have served as catalysts for addressing gender inequalities and advocating for the fulfillment of women's daily needs across different spheres of society.

Due to historical and cultural reasons, Chinese feminist discourse has specific characteristics different from its Western counterpart. As South Asian feminist Kamla Bhasin remarked, ``\textit{Feminism is like water. It's everywhere but it takes the shape of the container into which it is poured. My feminism is different...because my patriarchy is different} \cite{bhasin1999feminism}.'' Factors such as cultural traditions, societal expectations, and historical influences in China have collectively contributed to a distinct framing of issues \cite{wu2019made}. Chinese women have long lived within the patriarchal framework of Confucian ethics, which are essentially masculine and promote male dominance \cite{han2018searching}. To this day, the traditional norm of ``men are breadwinners, and women are homemakers'' still influences women's choices between family and career in everyday life, while also impacting public opinions and perceptions of women \cite{chen2021gender,mingmo}. Besides, everyday feminism on social media in China has its opportunities and challenges. Social media platforms have become crucial avenues for ordinary users to participate in feminist advocacy
\cite{pruchniewska2019everyday}. These platforms provide an opportunity for women to create their own narratives, and have their voices genuinely heard and shared
\cite{mao2020feminist}. Despite the presence of opportunities, there are also accompanying challenges. One of the challenges is the stigmatization of feminism, which often fuels gender debate on social media \cite{yang2023rural}. Feminism in China is often criticized as rural feminism (田园女权) that promotes male hatred and advocates for women's rights without responsibilities \cite{mao2020feminist,yin2021intersectional}. This stigma makes it difficult for feminists advocating for genuine equality to escape the association with pastoral feminism, and some refuse or even fear to identify themselves as feminists \cite{peng2020feminist,yang2022research}. Under such a socio-cultural context, gender debate in China emerges with contrasting viewpoints. Two representative cases demonstrating this phenomenon are described in Section \ref{cases}.


Gender debate that arises from everyday feminism on social media in China has not received sufficient investigation. This work aims to fill this gap by providing a comprehensive understanding of how public gender debate unfolds within the social and cultural context of China.

\subsection{Gender Debate on Social Media}

In recent years, gender issues have become prominent topics on Chinese social media \cite{peng2020feminist,zhang2023reflections,jiang2022swsr}. Weibo, one of China's largest social platforms, has emerged as a battleground for gender debate \cite{yutong2023hostility}. Confronted with challenges in organizing offline events, feminist groups have resorted to online campaigns to advocate for women's needs and address issues such as insufficient public infrastructure \cite{rw1}, sexual harassment \cite{rw2} and domestic violence \cite{rw3}. In response to the rising concerns of conflict-inciting content, Weibo introduced a mechanism for users to report instances of ``provoking hate'', such as sexism and racism \cite{rw4}. It deleted 54 accounts and temporarily muted 472 users for engaging in gender opposition, hate speech and conflict incitement on December 7, 2021 \footnote{https://weibo.com/1934183965/L4WmlfNub?pagetype=profilefeed}.

In the realm of public debate, \textit{incivility} and \textit{constructiveness} are two critical narrative aspects in HCI and CSCW \cite{masrani2023slowing,cheng2017anyone,baughan2021someone}. \textit{Incivility} on social media is defined as an act of sending or posting mean text messages intended to mentally hurt, embarrass or humiliate another person \cite{maity2018opinion}. This type of behavior could have detrimental effects, including the depletion of the active user base \cite{maity2018opinion}, the damage to the credibility of media outlets \cite{anderson2018toxic}, and the diminishing open-mindedness \cite{borah2014does}. Conversely, promoting \textit{constructive communication} is of utmost importance for fostering a healthy public discourse, and it is widely acknowledged as the ideal approach for engaging in online discussions \cite{coe2014online, sapiro1999considering,jaidka2019brevity,baughan2021someone,cheng2017anyone}. \textit{Constructive communication} entails clear and effective expression, politeness, and the provision of justifications for one's viewpoints \cite{towne2012design}. In gendered discourse, a plethora of research primarily focused on the uncivil aspect \cite{bou2014introduction,bou2014conflict,anderson2014public}. For example, gendertrolling, a phenomenon characterized by multiple individuals using vicious language to insult a specific gender, posed a significant threat to targeted individuals \cite{mantilla2013gendertrolling}. The users of an online discussion forum Incel, a virtual community of isolated men without a sexual life, viewed women as the source of their issues and usually utilized the forum to express misogynistic hate speech \cite{jaki2019online}. Nevertheless, the constructiveness attribute of gender debate has received relatively less attention in discussions surrounding gender issues.

Another important perspective of online debates is the \textit{stance} of users. For example, Baughan et al. examined how political identity, such as liberal and conservative accounts on Twitter, affects civility during political disagreements \cite{baughan2022shame}. Rho et al. highlighted the presence of in-group and out-group dynamics among commenters on Breitbart (representing a right-wing viewpoint) and DemNow (representing a far-left viewpoint) regarding the \#MeToo movement \cite{rho2018fostering}. Breitbart commenters engage in derogatory language towards \#MeToo participants reflecting out-group derogation, while DemNow commenters exhibit in-group favoritism based on race and socioeconomic factors. These biases have the potential to polarize the movement and undermine its initial solidarity. Additionally, in-group bias, a phenomenon in which people respond more favorably to those with whom they share a group identity \cite{baughan2022shame}, causes individuals to react more positively and constructively when faced with conflicts involving members of their own group, thereby promoting a more harmonious atmosphere for respectful disagreements within the group \cite{wenzel2008retributive,koval2012our}. 

Thus, our work delves into both incivility and constructiveness in the context of gender debate on Chinese social media, and explores the various stances while advocating women's daily needs. By analyzing gender discussions, we aim to contribute to a better understanding of the challenges and opportunities for fostering constructive dialogue surrounding gender issues on social media in China.

\subsection{Discursive Strategies in Gender Debate}

Discursive strategies are of great importance in online debate, given their potential to influence persuasive arguments \cite{tan2016winning}, impact how users evaluate and respond to content on social media \cite{cheng2017anyone, cavazza2014swearing,ismail2020defying,bkaczkowska2021you} and reveal social identities and perspectives \cite{rho2018fostering,anderson2014public}. 

In the context of gender debate, there has been considerable research focused on attacking strategies against women \cite{papadamou2021over,sobieraj2018bitch}. Users often employ specific linguistic strategies, such as the use of body parts, dehumanization representations, and other gendered terms to insult women \cite{sagredos2022slut}. For example, Goetz et al. conducted the first exploration into the various types of verbal insults used by men against their intimate female partners, encompassing derogatory comments like women's attractiveness and their value as a person \cite{goetz2006adding}. Herring et al. investigated strategies used by a ``troller'' attempting to disrupt a feminist web-based discussion forum \cite{herring2002searching}. 

Nevertheless, gender debate observed in this work exhibits mutual interaction, highlighting the importance of considering both sides in gender debate. Moreover, many works examining strategies in gendered discourse were theory-driven, e.g., how Discourse-Historical Approach (DHA) \cite{reisigl2017discourse} was applied to argumentation strategies in an online male separatist community \cite{aiston2022argumentation}. However, what are the bottom-up strategies developed by users in gender debate, and what are the effects of these strategies, are still largely under-explored. This work contributes to exploring a comprehensive taxonomy of debating strategies in gender debate (RQ1) and understanding how they influence user participation and user response (RQ2).

\section{METHOD}

\begin{figure*}
	\centering
		\includegraphics[scale=0.6755]{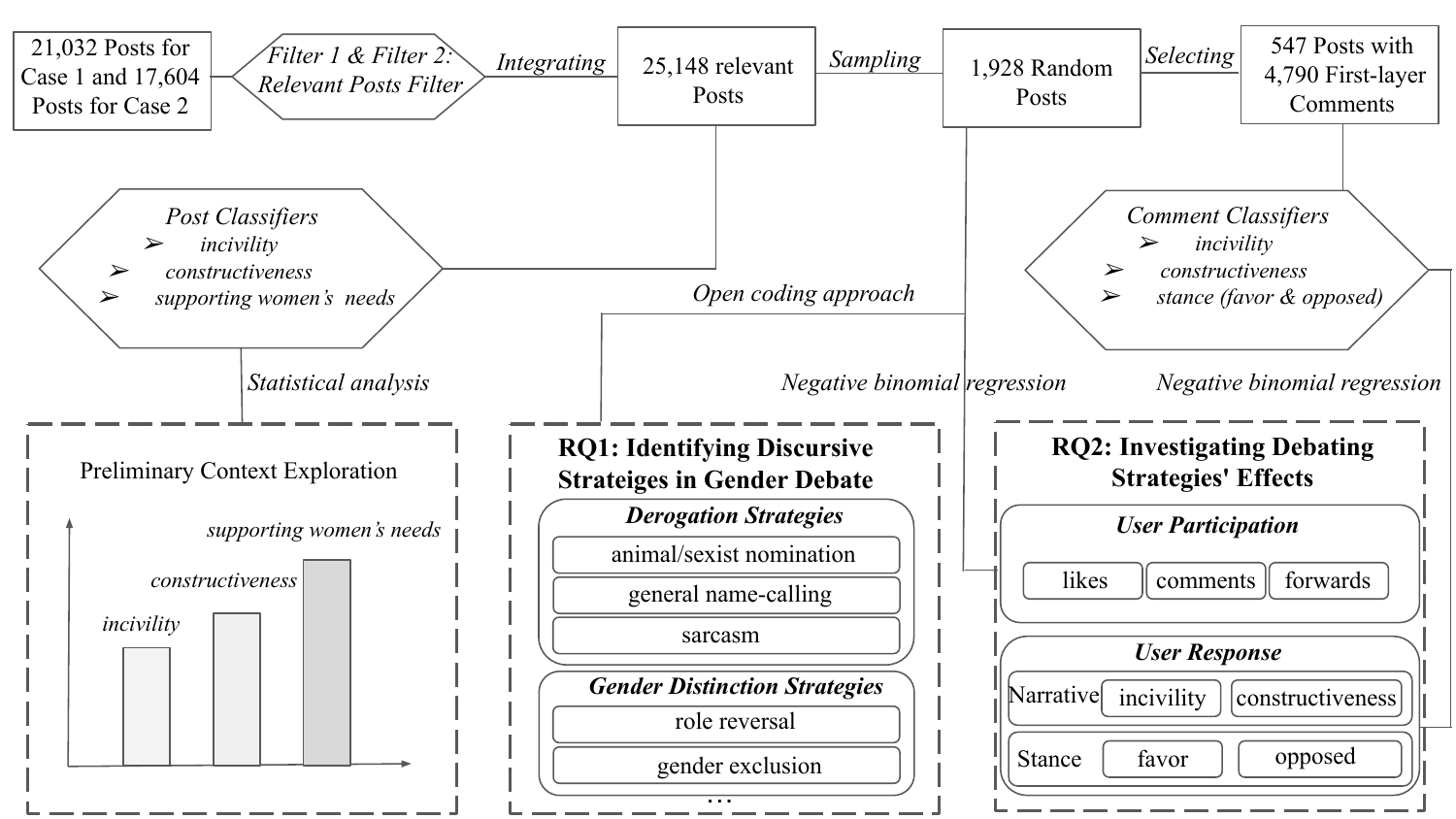}
	\caption{The analytical flow to understand discursive strategies in the gender debate.}
        \Description{The analytical flow of our work. We have 21,032 posts for Case 1 and 17,604 posts for Case 2. After filtering, we obtain 25,148 relevant posts. Based on these data, we classify these posts according to their attributes including incivility, constructiveness and supporting women's needs, and conduct statistical analysis for these attributes to get a preliminary context exploration. Then we sample 1,928 random posts to identify discursive strategies in gender debate through open coding approach (R1). After that, through negative binomial regression, we investigate debating strategies' effects on user participation based on 1,928 posts, and user response based on 547 posts within 4,790 comments (R2).}
	\label{FIG: method}
\end{figure*}

This section describes the mixed-methods approach employed to understand discursive strategies in gender debate on social media. First, we introduce two representative cases of gender debate within everyday feminism that we focus on in this study in Section \ref{cases}. Subsequently, Section \ref{data} describes data collection and pre-processing procedures. In Section \ref{context}, we delve into the preliminary exploration of the contextual factors. Next, we present the methodology employed to address our research questions. Specifically, in Section \ref{method: strategies}, we explain how we adopt an open coding approach to identify and categorize the strategies in gender debate (RQ1). Then, in Section \ref{effects}, we utilize regression analysis to examine the impact of debating strategies on user participation and response (RQ2). The overall analytical flow is shown in Figure \ref{FIG: method}.

\subsection{The Study Context: Two Representative Cases of Gender Debate within Everyday Feminism in China}\label{cases}

When speaking out women's daily needs on social media in China, there exists a heated gender debate. We describe two representative cases of gender debate during feminist advocacy as the context in this study. The platform this work is situated in is Weibo, one of the biggest social media platforms in China~\cite{chen2021exploring,qu2011microblogging}.

The first case is \textit{the demand for selling menstrual products on high-speed railway}. A gender debate erupted after a female passenger expressed her frustration online about the lack of menstrual products on high-speed railway when she unexpectedly experienced her period \footnote{\url{https://www.scmp.com/news/people-culture/trending-china/article/3193156/sorry-only-snacks-and-souvenirs-china-railways}}. The phenomenon gained intense debate on Weibo with more than 830 million views and 200,000 discussions (including original posts and retweets), with the emergence of a large volume of conflicting viewpoints \cite{case1.1}. For example, while some expressed surprise at the absence of women's sanitary products on trains and called for operators to provide them, others argued that it might be unreasonable since only a small proportion of passengers would require them.

The second case is \textit{the demand for sufficient female restrooms}. Women frequently encountered long queues and delays when waiting for access to overcrowded restrooms, but men enjoyed quicker and more convenient restroom experiences \footnote{https://www.globaltimes.cn/page/202112/1243127.shtml}. The gender debate took place as several users supported increasing the number of female toilets, while others thought that it unfairly disadvantaged men, with more than 420 million views and 53,000 discussions on Weibo \cite{case2.1, case2.2, case2.3}. These two cases exemplify the everyday challenges faced by women in China, which provide a lens to understand gender debate within everyday feminism and discursive strategies in the process.

\subsection{Data Collection and Pre-processing}
\label{data}
\subsubsection{Data Collection}
\label{data collection}
Prior to data collection, we browsed posts related to the two representative cases on Weibo. We observed that some users might attach specific hashtags (e.g., ``\#A woman claims she couldn't buy menstrual products while on high-speed railway during her period\#/\#女子称高铁上来例假买不到卫生巾\#'') when discussing these cases, while others might only use plain text for relevant discussion that contain specific keywords characterizing the case (e.g., ``menstrual products on high-speed railway/高铁卫生巾''). Consequently, to comprehensively collect data, we selected a keyword-based method to retrieve posts which could contain both situations. The selection of keywords aimed to encompass as many relevant posts as possible while excluding irrelevant ones. Initially, we searched for posts with the most direct phrases associated with the two cases (i.e., ``menstrual products on high-speed railway/高铁卫生巾'') and ``more women's toilets/多女厕''). Then, to expand the keyword set, we examined these searched posts to identify alternative expressions frequently appearing in the posts that characterized the cases, and repeated the search and identification steps based on the new keyword. This iterative process continued until no new keywords were discovered. For Case 1, we identified ``menstrual products on high-speed railway/高铁卫生巾'' as the keyword without expansion as it could well capture the case. For Case 2, we recognized ``more women's toilets/多女厕'', ``increase women's toilets/增加女厕'', and ``convert to women's toilets/改女厕'', all highlighting the need for increased female restrooms. After that, we employed WeiboSuperSpider tool \cite{WeiboSuperSpider} to retrieve all posts containing the identified keywords. In order to ensure comprehensive coverage of the data for these two cases, we collected posts with corresponding comments and user information on Weibo from January 1, 2013, to July 5, 2023, spanning a period of ten years, with the assistance of WeiboSuperSpider tool \cite{WeiboSuperSpider}. In total, there were 21,032 original posts, 81,822 comments and 18,848 distinct users for Case 1, and 17,604 original posts, 105,717 comments and 13,411 distinct users for Case 2. There was an overlap of 564 users between the two datasets.



\subsubsection{Data Pre-processing}
\label{processing}
Although we carefully selected suitable keywords, it was inevitable that the collected data might include some noise that was irrelevant to the intended scenario, such as advertising. To promote the purity of the dataset for further analysis, we implemented two filters (Filter 1: filtering relevant posts for Case 1, and Filter 2: filtering relevant posts for Case 2 as shown in Figure \ref{FIG: method}). They were two binary text classifiers designed to determine whether posts were related to this context. To construct these filters, two authors independently coded a random sample of 100 posts for each case on their relevance. Cohen's Kappa, a statistic for assessing inter-rater reliability in qualitative items, is generally considered as a more robust measure than percent agreement as it considers the likelihood of chance agreement \cite{mchugh2012interrater}. This round of coding established substantial inter-rater agreement (Filter 1: Cohen's Kappa = 1.00, agreement ratio = 100\%; Filter 2: Cohen's Kappa: 0.98, agreement ratio = 99\%). Then, the two coders discussed and reached a consensus on annotations. Subsequently, each \nobreak author individually annotated an additional 450 posts, resulting in a total of 1,000 posts as the training dataset. We tried traditional machine-learning methods (i.e., SVM \cite{cristianini2000introduction}and XGBoost \cite{chen2016xgboost}) and deep-learning methods (i.e., LSTM \cite{hochreiter1997long}, GRU \cite{cho2014learning}, and BERT \cite{devlin2018bert} fine-tuned with BERT-wwm \cite{cui2021pre}). To mitigate the overfitting of the deep-learning methods, we employed a Dropout layer with a parameter of 0.2 and utilized the Adam optimizer during the training. Among the evaluated models, BERT demonstrated the best performance for both Filter 1 (F1 score: 0.95) and Filter 2 (F1 score: 0.93). These models were then applied to filter all the collected posts, with 18,833 posts for Case 1 and 6,315 posts for Case 2.

During the preprocessing, we observed a significant homogeneity in the dataset, as both cases centered on advocating for women's needs and exhibited similar language patterns. Furthermore, there were occasional references to each other within posts, (e.g., \textit{``Just when we thought the drama with sanitary pads on high-speed railway was over, now we've got another one: universities deciding to convert men's restrooms into women's restrooms.''}). Given our goal was to examine gender debate stemming from feminist advocacy within everyday feminism rather than focusing on specific instances, we merged the data from both cases for further analysis.

\subsection{Preliminary Exploration on the Context}
\label{context}
Before exploring discursive strategies and their impacts on gender debate, we conducted a preliminary investigation of the discussion landscape, which helps better contextualize the further analysis. We looked through extensive posts and found that discourses in gender debate not only involved substantial \textit{uncivil} content but also contained many \textit{constructive} discussions. Moreover, users' stances could be categorized into either \textit{supporting women's needs} or not. Previous research also underscored the significance of considering incivility, constructiveness, and user stance as crucial dimensions in public debate online \cite{ferreira2021shut, jaidka2019brevity,baughan2022shame,rho2018fostering}. The definitions and examples of these attributes were shown in Table \ref{table0}. We assigned them as binary values. For \textit{supporting women's needs}, we set 1 to posts that support women's needs, while assigning 0 to other cases that do not support women's needs or remain neutral. This choice enabled a focused exploration of the contrasting impact between supporting and non-supporting women's needs within the context of gender debate when advocating women's needs.

\begin{table*}[htbp]
\caption{The definitions and examples of post attributes including \textit{incivility}, \textit{constructiveness} and \textit{supporting women's needs}.}
\label{table0}
\centering
\scalebox{1}{
\begin{tabular}{p{3.5cm}|p{5.5cm}p{5cm}}
   \toprule
    & Definition & Example\\
   \midrule
   Incivility\_post &the use of language that is intended to mentally hurt, embarrass or humiliate another person, typically including provocative, disrespectful and offensive language.& \textit{If you don't have a pussy, stay out of pussy business.} \\
   Constructiveness\_post &the act of providing informative content that enhances others' understanding and progress of the discussion, often through logical reasoning or evidence.& \textit{Not everyone has a regular menstrual cycle. Sometimes even with prior preparation, unexpected situations can still occur.}\\
   Supporting women's needs &In our context, supporting women's needs involves endorsing the sale of menstrual products, advocating for more women's restrooms and promoting the understanding of women's experiences.& \textit{This is awesome! No need to queue for the women's restroom in the Forbidden City. This is something that should be strongly promoted. Indeed, the number of women's restrooms should increase.}\\
   \bottomrule
\end{tabular}}
\end{table*}


To delve deeper into \textit{incivility}, \textit{constructiveness} and \textit{supporting women's needs} within gender debate, we built three context classifiers for them, and conducted a statistical analysis to assess the distribution of three attributes. An incivility classifier (Context Classifier 1) was constructed for \textit{incivility} determining whether a post was uncivil, a constructiveness classifier (Context Classifier 2) was built for \textit{constructiveness} determining whether a post was constructive, and a position classifier (Context Classifier 3) was developed to decide whether supporting women's needs or not. To achieve this, we first randomly selected 1,000 posts from each of the two case datasets. We obtained 1,928 posts from the whole dataset, after filtering the noise driven by the deep-learning classification in Section \ref{processing}. Then two authors coded 100 samples from the random subdataset (Context Classifier 1: Cohen's Kappa = 0.84, agreement ratio = 92.5\%; Context Classifier 2: Cohen's Kappa = 0.75, agreement ratio = 91\%; Context Classifier 3: Cohen's Kappa: 0.90, agreement ratio = 95\%). Subsequently, after reaching mutual consensus on these samples, each author individually annotated half of random subdataset, resulting in a total of 1,928 posts as the training datase for context classifiers. Based on the outstanding performance of BERT in Section \ref{processing}, BERT was selected for these tasks as well (F1 score of Context Classifier 1: 0.85, F1 score of Context Classifier 2: 0.82, F1 score of Context Classifier 3: 0.75). We also illustrated the temporal change of gender debate in Appendix \ref{Temporal}. 

\subsection{RQ1: Identifying Discursive Strategies in Gender Debate}
\label{method: strategies}
Drawn on Discourse-Historical Approach (DHA) \footnote{DHA as a classic Critical Discourse Analysis (CDA) approach concludes five types of discursive strategies and interprets discourse in its historical and cultural contexts \cite{reisigl2017discourse}, which is commonly employed in gender discourse analysis \cite{aiston2022argumentation,sagredos2022slut}.}, we employed an open coding approach \cite{flick2022introduction} to identify discursive strategies of users in gender debate. This inductive approach allowed the codes to emerge naturally from the analysis. Specifically, two authors initially conducted manual coding of 200 posts independently from a randomly selected dataset. We explored the strategies used by users from a nuanced perspective (saturation was achieved after coding nearly 100 posts). Subsequently, several rounds of meetings, comparisons, and discussions were held to consolidate similar strategies. Additionally, strategies exhibiting similar effects were grouped into broader categories, resulting in the hierarchical strategies (e.g., \textit{Derogation Strategies} includes \textit{animal/sexist nominations}, \textit{general name-calling}, and \textit{sarcasm}). These discursive strategies were described in detail in Section \ref{RQ1}, and the hierarchical structure of strategies was demonstrated in Table \ref{gender_table}.

In order to provide a quantitative description of strategies and prepare for regression analysis, it was necessary to determine which strategies were specifically employed in each post. This was accomplished using one-hot coding, where a label of 1 indicated the application of a particular strategy and 0 otherwise. During preliminary exploration in Section \ref{context}, we observed that a large number of posts contained implicit insults in Chinese, and often a combination of uncivil and constructive content coexisted within a single post. As a result, scaling up the analysis to the entire dataset using machine-learning methods could yield inaccurate results. To gain a more realistic understanding of the scenario, we employed a sampling approach. Specifically, we utilized the subdataset of 1,928 posts in Section \ref{context} which is randomly selected from the whole dataset. The coding process involved two coders evaluating whether each of the 1,928 posts applied the specified strategies. Specifically, to assess the inter-rater reliability, the coders initially re-coded 200 samples independently and compared their labels for each dimension. The results indicated an agreement ratio surpassing 90\% and Cohen's Kappa coefficient not less than 0.8 for every dimension. This demonstrated a significant level of agreement between the two coders. Following this, the coders proceeded to independently code half of the remaining posts.

Moreover, to explore the utilization of discursive strategies among different genders, we conducted a comparative analysis of gender-related differences in their usage patterns. Specifically, we used two proportions to gain insights into the gender representation of strategies: (1) \textit{Inter-gender proportion} refered to the proportion of male and female participation in posts that utilized a specific strategy. It focused on the overall usage ratio of \textbf{males and females} in a specific strategy; (2) \textit{Inner-gender proportion} measured the proportion of posts made by a specific gender that adopted a particular strategy, relative to the total number of posts made by that gender. It highlighted the prevalence of a specific strategy within \textbf{a particular gender}.  

\subsection{RQ2: Investigating Debating Strategies' Effects on User Participation and User Response}
\label{effects}
In this section, we conducted further investigations to explore the effects of these user-developed strategies. In debate scenarios, user participation (e.g., engagement indexes \cite{zhang2021mediatization}) and response quality (e.g., the use of evidence \cite{coe2014online}) are two typical focuses from previous works \cite{jaidka2019brevity,towne2012design,borah2014does}. The former serves as an indicator of the level of interactivity in the discussion, and the latter assesses the meaningfulness and productivity of the discourse. Moreover, with the existence of incivility, ensuring the discussion quality becomes even more challenging \cite{theocharis2016bad}. Drawing on the existing research focus and our own contextual exploration, we aim to investigate how these strategies affect \textit{user participation} (i.e., the number of likes, comments, and forwards), and \textit{user response} derived from comments (i.e., \textit{narrative} including \textit{incivility} and \textit{constructiveness} of comments, and \textit{stance} involving whether commentors are in \textit{favor} of or \textit{opposed} to a post within gender debate). By delving into these dimensions, we seek to gain insights into how the employed debating strategies shape user participation and responses.

\subsubsection{Dependent Variables}
\begin{itemize}
\item \textbf{\textit{User Participation.}} To assess user participation, we utilized all the 1,928 available sample data. We employed engagement indexes (i.e., likes, comments, and forwards) for each post. These indexes served as quantitative measures to evaluate the level of active engagement by users in the gender debate.
\begin{itemize}
\item \textbf{\textit{Likes}} (count): The total number of user likes received by a post.
\item \textbf{\textit{Comments}} (count): The total number of user comments obtained by a post.
\item \textbf{\textit{Forwards}} (count): The total number of times a post was forwarded by users.
\end{itemize}

\item \textbf{\textit{User Response.}} To investigate user response, we selected 547 posts with comments from all sample data, which collectively contained 4,790 first-layer comments. Given that comments have multiple layers which might lead to mutual influence of comments, we only considered the first layer of comments. We explored user response from the perspectives of \textit{narrative} and \textit{stance}. Specifically, in terms of \textit{narrative}, we constructed two binary text classifiers for \textit{incivility} and \textit{constructiveness} of comments, and built a pairwise text classifier for \textit{supporting} or \textit{opposing} commented posts.

\begin{itemize}
\item \textbf{\textit{Incivility\_comment}} (count): The number of comments exhibiting incivility under a post.

\item \textbf{\textit{Constructiveness\_comment}} (count): The number of comments containing constructive information under a post.

\item \textbf{\textit{Stance\_favor}} (count): The total number of comments in favor of a post.

\item \textbf{\textit{Stance\_opposed}} (count): The total number of comments opposing a post.

\end{itemize}

The incivility\_comment classifier (Comment Classifier 1) responsible for identifying uncivil comments, and the constructiveness\_comment classifier (Comment Classifier 2) designed to identify constructive comments, followed a similar building process to that of Filter 1 and Filter 2 in Section \ref{processing} (Comment Classifier 1: Cohen's Kappa = 0.92, agreement ratio = 96\%, F1 score = 0.79; Comment Classifier 2: Cohen's Kappa = 0.89, agreement ratio = 98\%, F1 score = 0.80).

Since \textit{stance\_favor} and \textit{stance\_opposed} were related to the corresponding posts, we developed a pairwise text classifier (Comment Classifier 3) to identify the semantic relations for post-comment pairs (i.e., favor, opposed and unknown). Initially, we randomly selected 1000 post-comment pairs from the dataset. Two authors independently coded the first 100 samples to determine the relationship between the post and comment (Comment Classifier 3: Cohen's Kappa = 0.81; agreement ratio = 88\%). After several rounds of discussions to resolve discrepancies, two authors annotated an additional 450 pairs each, resulting in a total of 1000 labeled samples. Then we also chose BERT due to its strong performance and the benefit of BERT's base training task of next-sentence prediction, which supports fine-tuning for sequence pair classification \cite{devlin2018bert}. With the basic structure of the sequence
pair classification \cite{ostendorff2020pairwise}, we separated tokens from posts and comments with the [SEP] token, identified the three types with a mask (token\_type\_ids), and jointly fed them into the model (Comment Classifier 3: F1 score = 0.70).

\end{itemize}

\subsubsection{Independent Variables}

\begin{itemize}

\item \textbf{\textit{Debating strategies}} (binary): We took discursive strategies identified in Section \ref{method: strategies} as the independent variables to explore how they correlated with the various dimensions of user participation and user response. 
\item \textbf{\textit{Control variables}}: Considering other relevant factors might also have an impact on our scenario, we included control variables. (1) Post context: the context including \textit{incivility\_post} (binary),  \textit{constructiveness\_post} (binary) and \textit{supporting women's needs} (binary) from posts might affect user participation and response in comments. (2) Post characteristics: \textit{post length} (count) and the number of \textit{hashtags} (count) are latent factors that contribute to user engagement~\cite{gkikas2022text}. (3) Poster information: the number of followers (i.e., \textit{follower} (count)) and the number of users following the poster (i.e., \textit{following} (count)) reflecting the social networks of the poster may also influence this context.

\end{itemize}

\subsubsection{Regression Analysis}
Since the dependent variables were characterized by count data and there was a significant disparity between their mean and variance, we opted for negative binomial regression \cite{hilbe2011negative} to explore the correlation between discursive strategies and user participation as well as user response. In order to avoid poor estimation caused by highly correlated features, we assessed multicollinearity using the Variance Inflation Factor (VIF) \cite{akinwande2015variance} and tested the correlation coefficients among the independent variables. Prior to conducting the regression, we verified that all independent variables had a VIF value below 5 and correlation scores were below 0.6. 

\section{FINDINGS}
By employing a mixed-methods approach, this work enhances our understanding of discursive strategies utilized by users in gender debate on social media and their corresponding effects. In this section, we commence by describing the findings from the preliminary context exploration in Section \ref{Context Description}. Following that, we present a taxonomy of strategies developed by users within the gender debate, as discussed in Section \ref{RQ1} (RQ1). Moreover, we demonstrate the correlations between these strategies and user participation as well as user responses in Section \ref{RQ2} (RQ2).
\subsection{Context Description}
\label{Context Description}
\begin{figure}[htbp]
	\centering
	{\includegraphics[width=0.96\columnwidth]{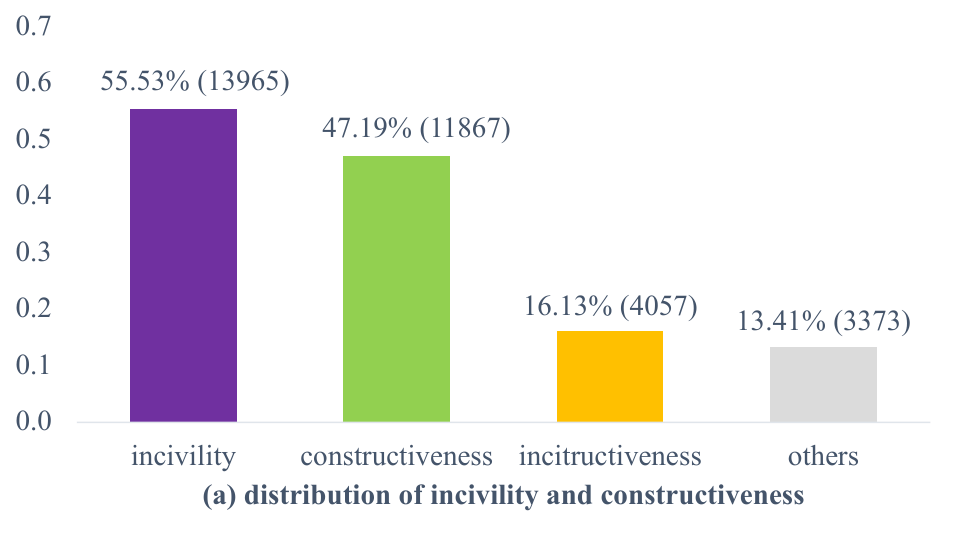}}\\{\includegraphics[width=0.96\columnwidth]{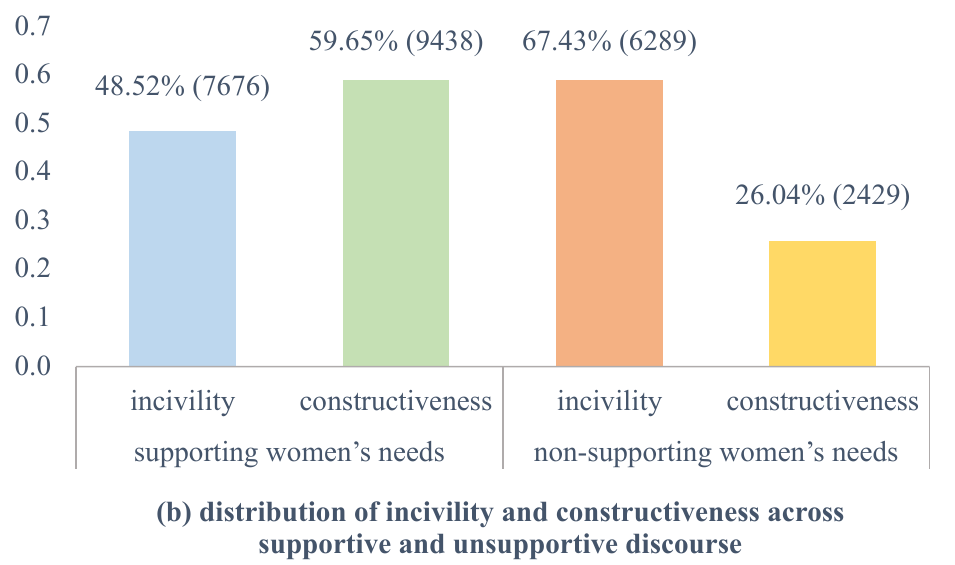}}\\
  \caption{The distribution of (a) uncivil and constructive content in gender debate, as well as (b) incivility and constructiveness within supportive and unsupportive discourse. ``Incitructiveness'' refers to posts that encompass both uncivil and constructive content.}
   \Description{The distribution of incivility and constructiveness in posts. Figure 2 (a) shows that 55.53\% (N=13965) posts contain uncivil content, 47.19\% (N=11867) posts include constructive content, and 16.13\% (N=4057) posts have both uncivil and constructive content. Figure 2 (b) shows that among posts supporting women's needs, there are 48.52\% (N=7676) uncivil posts and 59.65\% (N=9438) constructive posts, whereas among posts not supporting women's needs, there are 67.43\% (N=6289) uncivil posts and 26.04\% (N=2429) constructive posts.}
  \label{FIG: distribution}
\end{figure}

The results of the preliminary exploration were presented in Figure \ref{FIG: distribution}, revealing the following main findings: (1) The gender debate was heavily infused with incivility, with posts containing uncivil content accounting for 55.53\% of all posts. However, there were also a significant number of constructive contributions, constituting 47.19\% of the total. (2) Many posts exhibited both incivility and constructiveness, reaching a proportion of 16.13\%, indicating while some uncivil language was made, they were intertwined with meaningful and constructive content. (3) In posts advocating for women's needs, both incivility (48.52\%) and constructiveness (59.65\%) were prevalent. Additionally, there were more instances of constructiveness than incivility in these posts, contrary to the observations in posts not supporting women's needs. This suggested that individuals may statistically disseminate useful information to express their stance when speaking out for women.

\subsection{RQ1: Strategies in Gender Debate}
\label{RQ1}
In this section, we delved into the discursive strategies employed by users during gender debate on social media in Section \ref{findings: strategies} and gender-related difference of these strategies in Section \ref{gender difference}.
\subsubsection{Discursive Strategies}
\label{findings: strategies}
These strategies were organized into 5 distinct categories, encompassing a comprehensive set of 15 debating tactics. The percentage represented the proportion of posts that utilized a specific strategy relative to the overall number of posts.

\begin{itemize}

\item \textit{Derogation Strategies (40.61\%, N=783).}

The \textbf{\textit{Derogation}} category consisted of strategies primarily aimed at insulting or belittling a particular group. This category included three specific strategies: \textbf{\textit{animal/sexist nominations (7.26\%, N=140),}} \textbf{\textit{general name-calling (22.82\%, N=440),}} and \textbf{\textit{sarcasm (20.90\%, N=403).}} These strategies were used intentionally to undermine the value or status of the targeted group.

\textbf{\textit{Animal/sexist nominations}} compared individuals or groups to animals or employed sexist terms to degrade them. One user employed the analogy of infertile roosters to mock men who expressed opposition to the idea of purchasing sanitary pads on high-speed railway, implying that their concerns were baseless or irrational.

\begin{quote}
    \textit{``Seeing all the lengthy arguments and strong reactions, it's easy to think of infertile roosters who are upset about hens laying eggs and nesting. Cluck, cluck, cluck.''}
\end{quote}

Another user used a sexist term to warn men against joining the discussion.
\begin{quote}
    \textit{``Don't get involved with pussies if you don't have a pussy.''}
\end{quote}

\textbf{\textit{General name-calling}} referred to utilizing uncivilized language to directly insult or derogate others. For instance, one poster strongly criticized those who disagreed with the idea of selling sanitary pads.

\begin{quote}
    \textit{``I equally insult anyone who thinks it shouldn't sell sanitary napkins on the high-speed rail, go to hell, get hit by a car, or go climbing alone and fall off the cliff.''}
\end{quote}

\textbf{\textit{Sarcasm}}, described as ``yin yang guai qi'' (original Chinese: 阴阳怪气), was characterized by an abnormal and indistinguishable tone, which implied indirect expression of attitudes \cite{tian2022non}. The strategy was quite prevalent, with sarcasm being employed in nearly 1 out of 5 posts. Its most apparent manifestation was conveying the opposite of the intended meaning. In the context of converting male restrooms into female restrooms, a male sarcastically stated that men have no rights in this matter.
\begin{quote}
    \textit{``How can males even be considered human beings?''}
\end{quote}

Another form of \textit{sarcasm} was the use of symbols (e.g., ???) to indirectly express negative emotions (e.g., speechlessness or anger), or utilizing emojis (e.g., an upside-down smiling face) to convey mockery. For example, a poster expressed speechlessness and shock in response to a friend's viewpoint, so as to sarcastically criticize that perspective.

\begin{quote}
    \textit{``My female colleague said, `Sanitary pads are personal items, so why should they be sold on high-speed trains?' and I'm like, `???'''}
\end{quote}


\item \textit{Gender Distinction Strategies (9.39\%, N=181)}.

\textbf{\textit{Gender Distinction}} focused on strategies that highlighted the distinctions between genders. Strategies under this category were: \textbf{\textit{role reversal (2.02\%, N=39)}} and \textbf{\textit{gender exclusion (7.47\%, N=144).}} Despite their relatively lower proportions, this does not diminish the significance of these strategies.

\textbf{\textit{Role reversal}} entailed the deliberate act of exchanging or shifting traditional gender roles and expectations, in order to compare experiences between genders and draw attention to the inherent disparities. It often employed techniques of exaggeration and contrast.

Exaggeration was used by presenting implausible scenarios to underscore physiological disparities between genders or the unfairness of societal expectations, such as:
\begin{quote}
    \textit{``I hope males get their periods
for 32 days a month..''}
\end{quote}

Furthermore, the use of contrasting rhetoric served to reveal the injustices and inequalities brought about by the reversal of gender roles, such as:

\begin{quote}
    \textit{``Trains have smoking rooms for males in every compartment... But now there is fierce opposition to selling sanitary pads on high-speed trains. This world is indeed designed to cater to males.''}
\end{quote}

\textbf{\textit{Gender exclusion}} aimed at excluding individuals of a specific gender from participating in or commenting on issues related to another gender, highlighting gender differences and power dynamics, like:

\begin{quote}
    \textit{``I don't understand why some influential men always feel the need to comment on women's issues. How does this concern you? It's none of your business!''}
\end{quote} 

\item \textit{Intensification Strategies (35.53\%, N=685).}

\textbf{\textit{Intensification}} was used to fuel or escalate a discussion or argument, with the goal of amplifying the impact and significance of a viewpoint. It contained \textbf{\textit{recontextualization (22.30\%, N=430),}} \textbf{\textit{demand escalation (2.28\%, N=44),}} and \textbf{\textit{overgeneralization (18.57\%, N=358).}}

\textbf{\textit{Recontextualisation}} strategy involved quoting or paraphrasing ideas from the opposing side (often the outrageous and extreme opinions) and then refuting or ridiculing them. This strategy sought to undermine the opposing viewpoint by highlighting its extreme or flawed aspects. A woman quoted a comment from a man to express her anger: 

\begin{quote}
    \textit{``...I saw a comment from a guy: `If they can sell sanitary pads, they can sell condoms too! I really hope for real-name registration to see if these people are just animals!'''}
\end{quote} 

\textbf{\textit{Demand escalation}} was the act of escalating or amplifying gender-specific needs or demands. For example, a user put forward an additional demand that pushed the boundaries of the original request.

\begin{quote}
    \textit{``The high-speed railway should not sell the menstruation pads, but should give them for free!''}
\end{quote} 

\textbf{\textit{Overgeneralization}} extended conclusions or generalized a specific issue to broader societal concerns, involving the discussion of other significant gender-related social problems that may not be directly related to the initial topic. In this case, a male elevated the event of converting men's restrooms to women's restrooms to the level of class conflicts and claimed that women had a conspiracy to control men.

\begin{quote}
    \textit{``\#Sichuan University Responds To Conversion Of Men's Restrooms To Women's\# Why does society seem to allow gender hostility? Because it can mask class conflicts. Why is there an emphasis on male dominance? Because societal education of males is designed to make it easier for females to control them.''}
\end{quote} 
\item \textit{Mitigation Strategies (18.83\%, N=363).}

\textbf{\textit{Mitigation}} tended to alleviate the intensity of gender debate or avoid unnecessary misunderstandings, consisting of \textbf{\textit{suggestion (13.07\%, N=252),}} \textbf{\textit{self-defense (0.78\%, N=15),}} and \textbf{\textit{gender perspective-taking (5.39\%, N=104).}}

\textbf{\textit{Suggestion}} could be described as providing advice or recommendations regarding the solutions to the gender debate. One form suggesting concrete measures, involving specific actions or steps to address the issue, such as:

\begin{quote}
    \textit{``It's necessary to have sanitary pads available (like hanging a vending box in the restroom) because not everyone remembers or carries them with them all the time.''}
\end{quote} 

Another form was behavioral or ideological guidance, such as maintaining a particular mindset. For instance, one user advised that people should express their thoughts and let others listen to their challenges.

\begin{quote}
    \textit{``We should consider expressing ourselves and sharing our thoughts. If our perspectives are not actively taken into consideration by the world, it becomes crucial for us to communicate and convey them.''}
\end{quote} 

\textbf{\textit{Self-defense}} was a strategic approach where individuals expressed their viewpoints while encouraging others to avoid excessive criticism or personal attacks. It usually incorporated qualifying phrases or terms to mitigate potential negative responses and address anticipated criticism. A common manifestation of this strategy was the inclusion of parentheses at the beginning or end of a post. For instance, a poster kindly asked others to respond respectfully.

\begin{quote}
    \textit{``...If they sell it for thirty yuan on the train, would you find it too expensive and choose not to buy, then directly ask the help of other passengers... (This is my personal opinion, so please be respectful if you have different views).''}
\end{quote} 

A user limited the insulting target to a certain group of males:

\begin{quote}
    \textit{``...Do we need men to point fingers and make trouble? Do you experience menstruation??? (Clarification: I'm just referring to some idle men who have nothing better to do but cause trouble.)''}
\end{quote} 

\textbf{\textit{Gender perspective-taking}} referred to adopting or expressing viewpoints that support the opposite gender, such as:

\begin{quote}
    \textit{``A university converts men's toilets to women's toilets...A male student said it makes sense since there are more female students on campus.''}
\end{quote} 

\item 
\textit{Cognizance Guidance Strategies (66.18\%, N=1276).}

\textbf{\textit{Cognizance Guidance}} intended to raise public awareness of the significance and challenges pertaining to gendered issues. This category included \textbf{\textit{sympathy invoking (23.76\%, N=458),}} \textbf{\textit{evidence informing (47.87\%, N=923),}} \textbf{\textit{gender-related educating (12.24\%, N=236),}} and \textbf{\textit{(potential) problem raising (9.75\%, N=188).}}

\textbf{\textit{Sympathy invoking}} presented gender-specific personal experiences to raise empathy, compassion, or understanding. One woman shared her terrible experience: 

\begin{quote}
    \textit{``I once took the high-speed railway but didn't bring sanitary pads. My whole body felt very painful, but I had to take the train for five hours without sanitary pads. It was really painful. Can't they pay more attention to the embarrassment and needs of women?''}
\end{quote} 

\textbf{\textit{Evidence informing}}, as the most commonly employed strategy, utilized various types of evidence, including factual truths, cultural norms and general operational mechanisms to strengthen one's argument. Some users provided factual observations and phenomena on trains:\begin{quote}\textit{``On the high-speed trains in Yunnan province, there are sanitary pads available for purchase, and they have a high sales volume.''}
\end{quote}~\begin{quote}\textit{``On trains, they can sell beer, peanuts and other non-essential items, but they cannot sell sanitary pads which are essentially necessary for women.''}
\end{quote} 

Another user shared insights into cultural norms surrounding menstruation:

\begin{quote}
    \textit{``From a young age, what we were taught was that `menstruation is shameful', that menstrual blood is dirty and inauspicious. According to the customs in my hometown, during menstruation girls are not allowed to offer incense.''}
\end{quote} 

One poster mentioned operational mechanisms of marketing in the following post:

\begin{quote}
    \textit{``...1. Where there is demand, there will be supply. 2. The reason for the lack of supply is insufficient demand, which prevents the formation of a stable market on high-speed trains..., Conclusion: The decision to sell sanitary pads should strictly follow market rules, and they should be sold as much as possible...''}
\end{quote}

\textbf{\textit{Gender-related educating}} encompassed the practice of educating others by providing descriptions and explanations of gender-related knowledge. For instance, one user disseminated essential information about women's physiological and hygiene knowledge to the community.

\begin{quote}
    \textit{``The timing of a woman's period, whether it comes early, late, or aligns with a period tracking app, is not something women can control.''}
\end{quote}

\textbf{\textit{(Potential) problem raising}} highlighted (potential) issues, challenges, or consequences that might arise from addressing gender-specific demands. One poster complained about the dissatisfaction with the measures implemented:

\begin{quote}
    \textit{``...The issue of long queues in the women's restroom has been resolved, but now even the men's restroom requires queuing! Instead of converting men's restrooms into women's restrooms, it would be better to construct an additional women's restroom.''}
\end{quote}

\end{itemize}

\subsubsection{Gender-related Difference of Strategies in Gender Debate}
\label{gender difference}
Table \ref{gender_table} demonstrated the differences of discursive strategies between genders. The comparison yielded some important findings: (1) Females (58.04\%, N=1,119) were more active than males (39.57\%, N=763) in gender debate in the context of supporting women's needs. Additionally, there were 41 posts where gender information was not disclosed; (2) Females might be more likely to use \textit{general name-calling}, \textit{sarcasm}, and \textit{gender exclusion} strategies than males due to higher inner-gender proportion; (3) Despite the lower overall participation of males, they displayed a higher tendency to utilize strategies such as \textit{suggestions}, \textit{evidence informing}, and \textit{(potential) problem raising} compared to females.


\begin{table*}[htbp]
	\footnotesize
	\centering
	\caption{Statistical results of gender-related differences of strategies in gender debate. n refers to the number of posts, inter\_p is short for inter-gender proportion and inner\_p is an abbreviation for inner-gender proportion. To assess the significance of differences in strategies and context between genders, we utilized Fisher exact test, which is suitable for relatively small dataset (N=1000). *** p<0.001; ** p<0.01; * p<0.05.}
 \label{gender_table}
 \scalebox{1.2}{
\begin{tabular}{p{2cm}p{3cm}p{0.8cm}p{1cm}p{1cm}|p{0.8cm}p{1cm}p{1cm}}
\hline
\multicolumn{2}{l}{\multirow{2}{*}{}}                            & \multicolumn{3}{c|}{Female (58.04\%, N = 1,119)}                       & \multicolumn{3}{c}{Male (39.57\%, N = 763)}                         \\
\cline{3-8}
\multicolumn{2}{l}{}                                             & n & inter\_p & inner\_p & n & inter\_p & inner\_p \\
\hline
\multicolumn{8}{l}{\textbf{Strategies}}                                                                                                                                         \\
\hline
\multirow{3}{*}{Derogation}         & animal/sexist nomination**   & 98     & 71.01\%             & 8.76\%            & 40     & 28.99\%             & 5.24\%            \\
                                    & general name-calling***       & 309    & 72.20\%             & 27.61\%           & 119    & 27.80\%             & 15.60\%           \\
                                    & sarcasm***                    & 276    & 70.23\%             & 24.66\%           & 117    & 29.77\%             & 15.33\%           \\
                                    \cline{1-2}
\multirow{2}{*}{Gender Distinction} & role reversal              & 22     & 57.89\%             & 1.97\%            & 16     & 42.11\%             & 2.10\%            \\
                                    & gender exclusion***           & 121    & 86.43\%             & 10.81\%           & 19     & 13.57\%             & 2.49\%            \\
                                    \cline{1-2}
\multirow{3}{*}{Intensification}    & recontextualisation        & 234    & 55.45\%             & 20.91\%           & 188    & 44.55\%             & 24.64\%           \\
                                    & demand escalation          & 24     & 54.55\%             & 2.14\%            & 20     & 45.45\%             & 2.62\%            \\
                                    & overgeneralization         & 213    & 62.28\%             & 19.03\%           & 129    & 37.72\%             & 16.91\%           \\
                                    \cline{1-2}
\multirow{3}{*}{Mitigation}         & suggestion***                 & 120    & 48.98\%             & 10.72\%           & 125    & 51.02\%             & 16.38\%           \\
                                    & self-defense               & 11     & 73.33\%             & 0.98\%            & 4      & 26.67\%             & 0.52\%            \\
                                    & gender perspective-taking***  & 28     & 27.45\%             & 2.50\%            & 74     & 72.55\%             & 9.70\%            \\
                                    \cline{1-2}
\multirow{4}{*}{Cognizance Guidance} & sympathy invoking          & 280    & 62.08\%             & 25.02\%           & 171    & 37.92\%             & 22.41\%           \\
                                    & evidence informing***         & 442    & 48.95\%             & 39.50\%           & 461    & 51.05\%             & 60.42\%           \\
                                    & gender-related educating   & 127    & 55.22\%             & 11.35\%           & 103    & 44.78\%             & 13.50\%           \\
                                    & (potential) problem rasing*** & 83     & 45.11\%             & 7.42\%            & 101    & 54.89\%             & 13.24\%           \\
                                    \hline
\multicolumn{8}{l}{\textbf{Context}}                                                                                                                                   \\
   \hline
\multirow{3}{*}{Post Content}       & opinion (supporting women's needs)                    & 731    & 58.90\%             & 65.33\%           & 510    & 41.10\%             & 66.84\%           \\
                                    & incivility***                 & 589    & 70.37\%             & 52.64\%           & 248    & 29.63\%             & 32.50\%           \\
                                    & constructiveness***           & 625    & 50.73\%             & 55.85\%           & 607    & 49.27\%             & 79.55\%  \\
                                    \hline
\end{tabular}}
\end{table*}

\subsection{RQ2: Effects of Strategies in Gender Debate}
\label{RQ2}
\subsubsection{Descriptive Statistics}
We provided a comprehensive explanation of discursive strategies and their differences between genders in Section \ref{RQ1}. Building upon that, this section focuses on dependent variables and control variables to establish a solid foundation for regression analysis.

Both dependent variables for user participation (i.e., \textit{likes}, \textit{comments}, and \textit{forwards}) and user response (i.e., \textit{narrative} and \textit{stance} of comments) exhibited a long-tail effect. This highlighted the presence of a skewed distribution, where a few highly influential posts had a significant impact on the outcomes, while the majority of posts had comparatively lesser engagement or influence.

In terms of control variables regarding user participation, users used a median of 102 characteristics (mean = 137.5, std dev = 168.6) and included a median of 1 hashtag per post (mean = 0.7, std dev = 0.8). They had a median of 326.0 followers (mean = 826,047.8), and followed a median of 354.5 accounts (mean = 649.2). However, there was a significant disparity in users' social network metrics, with standard deviations of 4,255,169.0 for follower count and 1,126.8 for following count. The dataset also contained 865 (44.87\%) uncivil content, 1,253 (64.99\%) constructive posts, and 1,271 (65.92\%) posts advocating for women's rights. 

For user response, we discovered that the posts had a median length of 119 characters (mean = 164.5, std dev = 209.9) and contained a median of 1 hashtag per post (mean = 0.8, std dev = 0.8). The median follower count was 1286.0 (mean = 1,588,910.0), and the median number of accounts followed was 448.0 (mean = 747.6). Similar to user participation, there was considerable variation in the social network metrics of users, with standard deviations of 5,700,226.0 for the follower number and 908.7 for the following number. Within the dataset, we identified 233 (42.60\%) instances of uncivil content, 387 (70.75\%) constructive posts, and 366 (66.91\%) posts supporting women's needs.

\subsubsection{Effects on User Participation}
Table \ref{TAB: participation} showed the predicted results of negative binomial regression models for user participation in gender debate regarding likes (Model 1a), comments (Model 1b), and forwards (Model 1c). We conclude the following important findings: 

\begin{itemize}
    \item \textbf{Beware of Derogatory Language.}
Adopting derogatory language, particularly employing \textit{sarcasm}, had a significant negative impact on \textit{user participation}. The presence of \textit{sarcasm} was associated with a considerable decrease in engagement metrics, including 35\% fewer likes, 36\% fewer comments, and a substantial 77\% reduction in forwards. These findings underscored the detrimental effect of employing derogatory language, such as \textit{sarcasm} and \textit{general name-calling}, as a communication strategy, which created a hostile and unwelcoming atmosphere that hindered user engagement across various forms of interaction.

\item \textbf{The Power of Role Reversal.}
The implementation of \textit{role reversal} might contribute to a significant increase in forwards, with each additional instance of \textit{role reversal} integrated into posts having an approximate 7.27-fold boost in the number of forwards. It suggested that by challenging traditional gender norms and expectations, \textit{role reversal} may have the potential to resonate with users and encourage them to share the content with others.

\item \textbf{The Positive Influence of Recontextualisation and Overgeneralization.}
There was a strong positive association between \textit{recontextualization} and user participation in the gender debate, with a substantial increase of 143\% more likes and 182\% more forwards. Similarly, \textit{overgeneralization} showed a significant positive correlation, following a remarkable increase of 196\% more likes and 131\% more comments. 

\item \textbf{The Mixed Impact of Suggestion.} The effect of \textit{suggestion} on user participation exhibited diverse outcomes. While it positively enhanced the number of likes, there was a decline in the number of forwards. This demonstrated that \textit{suggestion} could effectively capture users' attention. However, its ability to stimulate content sharing appeared to be comparatively limited. 

\item \textbf{The Effectiveness of Evidence and Education.} Posts that provided \textit{evidence} or offered gender-related \textit{education} experienced significant boosts in user participation. The incorporation of informing evidence showed a strong correlation with a 37\% increase in likes and an astonishing surge of 451\% in forwards. Furthermore, posts offering gender-related education garnered a remarkable rise of 259\% more likes and 278\% increase in forwards. This highlighted the importance of presenting factual information and fostering understanding to captivate users' engagement.

\item \textbf{The Effect of Control Variables: Off-Course User Participation.}
\textit{Supporting women's needs} and \textit{providing constructive information} were found to have a negative impact on user participation. This indicated that the public's engagement might tend to focus more on discussions that employ strategies like \textit{overgeneralization} involving more serious social problems, or \textit{recontextualization} referencing others' extreme statements, rather than those genuinely advocate for women's needs or providing constructive information.

\end{itemize}

\begin{table*}[htbp]
	\footnotesize
	\centering
	\caption{Results of negative binomial regressions for user participation in gender debate. IRR (Incidence Rate Ratio) indicates the ratio change of the dependent variable when increasing an independent variable by one unit. ***p<0.001; **p<0.01; *p<0.05.}
\scalebox{1.18}{
	\begin{tabular}{p{2cm}p{3cm}p{0.9cm}p{0.9cm}p{0.9cm}p{0.9cm}p{0.9cm}p{0.9cm}}
		\hline
		& & \multicolumn{2}{l}{M1a: Like\_num}&\multicolumn{2}{l}{M1b: Comment\_num} & \multicolumn{2}{l}{M1c: Forward\_num}\\
		\cline{3-8}
		& &IRR&Std. Err.&IRR&Std. Err.&IRR&Std. Err.\\
		\hline
		
		\multicolumn{8}{l}{\textbf{Strategies}}\\
		\hline
		\multirow{3}{1in}{Derogation} 
	    &animal/sexist nomination&\gr0.84&\gr0.129&\gr0.91&\gr0.147&0.30**&0.197\\
		&general name-calling&0.67*&0.090&\gr0.90&\gr0.105&0.41**&0.131\\
	    &sarcasm&0.65*&0.091&0.64*&0.103&0.23***&0.132\\
        \cline{1-2}
		\multirow{2}{1in}{Gender Distinction} 
		&role reversal&\gr2.20&\gr0.202&\gr2.11&\gr0.219&7.27***&0.246\\
            &gender exclusion&\gr0.93&\gr0.129&\gr0.58&\gr0.155&\gr0.67&\gr0.188\\
		
		\cline{1-2}
		\multirow{3}{1in}{Intensification} 
	&recontextualisation&2.43***&0.077&\gr1.27&\gr0.086&2.82***&0.105\\
		&demand escalation&0.42*&0.213&\gr1.28&\gr0.228&\gr0.77&\gr0.295\\
    &overgeneralization&2.96***&0.082&2.31***&0.090&\gr1.55&\gr0.116\\
            \cline{1-2}
    	\multirow{3}{1in}{Mitigation}
		&suggestion&1.76**&0.102&\gr0.81&\gr0.108&0.42**&0.146\\
		&self-defense&\gr0.66&\gr0.334&\gr0.90&\gr0.372&\gr0.30&\gr0.580\\
		&gender perspective-taking&\gr0.87&\gr0.135&\gr1.60&\gr0.142&\gr0.70&\gr0.168\\	
  
		\cline{1-2}
		\multirow{4}{1in}{Cognizance Guidance}
		&sympathy invoking&\gr1.02&\gr0.080&\gr1.10&\gr0.084&\gr0.89&\gr0.104\\
		&evidence informing&1.37*&0.070&\gr1.17&\gr0.079&5.51***&0.096\\
		&gender-related educating&3.59***&0.099&1.93**&0.101&3.78***&0.124\\
		&(potential) problem raising&\gr0.75&\gr0.113&\gr0.78&\gr0.122&0.52*&0.153\\
		\hline
		\multicolumn{7}{l}{\textbf{Control Variables}}\\
            \hline
            \multirow{3}{1in}{Post Content} 
		&opinion (supporting women's needs)&0.52***&0.069&0.58***&0.078&0.65*&0.095\\
		&incivility&\gr1.17&\gr0.099&\gr0.94&\gr0.110&\gr1.36&\gr0.139\\
		&constructiveness&0.42***&0.080&\gr0.83&\gr0.092&0.31***&0.113\\

		\cline{1-2}
            \multirow{2}{1in}{Post Characteristics}
		&post length&1.00***&0.000&1.00***&0.000&1.00***&0.000\\
		&hashtag\_num&1.67***&0.043&1.28**&0.044&0.81*&0.052\\

		\cline{1-2}
		\multirow{2}{1in}{Poster Info}
		&follower&1.00***&0.000&1.00***&0.000&1.00***&0.000\\
		&following&1.00***&0.000&1.00***&0.000&1.00***&0.000\\
		\hline\hline
           (Intercept)&&3.86***&0.087&0.48***&0.097&0.25***&0.115\\
           No. Observations&&\multicolumn{2}{c}{1928}&\multicolumn{2}{c}{1928}&\multicolumn{2}{c}{1928}\\
           Pseudo R-squ. &&\multicolumn{2}{c}{0.2715}&\multicolumn{2}{c}{0.1754}&\multicolumn{2}{c}{0.2250}\\
           Log Likelihood&&\multicolumn{2}{c}{-3953.8}&\multicolumn{2}{c}{-2552.9}&\multicolumn{2}{c}{-1823.9}\\
            \hline
	\end{tabular}}
\label{TAB: participation}
\end{table*}

\subsubsection{Effects on User Response}
\label{4.3.3}

Table \ref{TAB: response} showed the predicting results of negative binomial regression models for a user response in gender debate regarding (1) narrative: uncivil discourse (Model 2a) and constructive discourse (Model 2b), and (2) stance: in favor of a post  (Model 2c) and opposed to a post (Model 2d). The key findings were as below: 

\begin{itemize}

\item \textbf{The Negative Influence of Animal/Sexist Nomination.} This strategy had a consistent negative impact on all attributes of responses, irrespective of whether the comments included incivility, provided constructive information, or expressed agreement or disagreement. The finding suggested that the use of animal/sexist nominations would reduce the likelihood of receiving responses.

\item \textbf{Sarcasm and Overgeneralization Might Contribute to Opposing Views.} \textit{Sarcasm} inclined to attract users with opposing viewpoints, within an increase of 241\% in opposing posts, while supporting posts decreased by 35\%. Similarly, \textit{overgeneralization} captured over 287\% opposing posts. This finding indicated that \textit{sarcasm} and \textit{overgeneralization} tended to elicit responses from those who disagree, potentially fostering polarization.

\item \textbf{Recontextualization Potentially Engages Discussions and Supporting Views.} The utilization of \textit{recontextualization} strategy demonstrated a correlation with a 187\% increase in uncivil posts and a 122\% increase in constructive posts. Additionally, it had attracted more supporting users, with a 107\% increase in supportive posts and a 48\% suppression of oppositional posts. This might be attributed to the nature of \textit{recontextualization}, which often involved referencing extreme attitudes that were completely contrary to one's own, making them easier to refute and attracting more in-group discussions and support.

\item \textbf{The Backfire Effect of Suggestion and Perspective-taking.} Although the intention of users adopting \textit{suggestion} and \textit{perspective-taking} was to solve problems and understand the challenges of other genders, it was related to an increase in disagreement. Specifically, \textit{suggestion} was associated with a 517\% increase in opposing posts and a 329\% increase in uncivil posts, while \textit{perspective-taking} had a 257\% increase in disagreement. 

\item \textbf{The Dual Impact of Strategies: Positive Correlation with Both Constructiveness and Incivility.} The strategies of \textit{sympathy invoking}, \textit{evidence informing}, \textit{gender educating}, \textit{(potential) problem raising} and \textit{suggestion} were positively correlated to constructive engagement or discussions among users from different stances. These strategies might serve as potential means to initiate valuable discussions. However, it was noteworthy that these strategies also had a simultaneous impact on incivility, following an increase in uncivil behavior within the discussions.

\end{itemize}

\begin{table*}[htbp]
	\footnotesize
	\centering
	\caption{Results of negative binomial regressions for user response in gender debate. IRR (Incidence Rate Ratio) indicates the ratio change of the dependent variable when increasing an independent variable by one unit. ***p<0.001; **p<0.01; *p<0.05 (Con refers to Constructiveness here).}
	\label{tab: reg_participation}
 \scalebox{1.13}{
	\begin{tabular}{p{2cm}p{2.9cm}p{0.46cm}p{0.9cm}p{0.46cm}p{0.9cm}|p{0.46cm}p{0.9cm}p{0.46cm}p{0.9cm}}
		\hline
		& & \multicolumn{4}{c|}{Narrative}&\multicolumn{4}{c}{Stance}\\
            \cline{3-10}
		& & \multicolumn{2}{l}{M2a: Incivility}&\multicolumn{2}{l|}{M2b: Con} & \multicolumn{2}{l}{M2c: Favor}& \multicolumn{2}{l}{M2d: Opposed}\\
		\cline{3-10}
		& &IRR&Std. Err.&IRR&Std. Err.&IRR&Std. Err.&IRR&Std. Err.\\
		\hline
		
		\multicolumn{8}{l}{\textbf{Strategies}}\\
		\hline
		\multirow{3}{1in}{Derogation} 
	    &animal/sexist nomination&0.46**&0.258&0.46**&0.276&0.43**&0.245&0.35**&0.387\\
		&name-calling&\gr1.06&\gr0.188&\gr1.05&\gr0.191&\gr1.35&\gr0.185&\gr0.82&\gr0.238\\
	    &sarcasm&0.70*&0.179&\gr0.80&\gr0.184&0.65*&0.171&3.41***&0.225\\
        \cline{1-2}
		\multirow{2}{1in}{Gender Distinction} 
		&role reversal&\gr0.97&\gr0.391&\gr1.13&\gr0.354&\gr1.66&\gr0.363&\gr1.40&\gr0.410\\
            &gender exclusion&\gr0.81&\gr0.300&\gr0.62&\gr0.325&\gr0.95&\gr0.289&\gr0.48&\gr0.473\\
		\cline{1-2}
		\multirow{3}{1in}{Intensification} 
	&recontextualisation&2.87***&0.155&2.22***&0.151&2.07***&0.154&0.52**&0.191\\
		&demand escalation&\gr0.65&\gr0.490&\gr0.69&\gr0.484&\gr0.46&\gr0.417&\gr0.99&\gr0.624\\
    &overgeneralization&1.71**&0.159&\gr1.21&\gr0.155&\gr0.79&\gr0.154&3.87***&0.187\\
            \cline{1-2}
    	\multirow{3}{1in}{Mitigation}
		&suggestion&4.29***&0.219&5.08***&0.207&2.51***&0.226&6.17***&0.230\\
		&self-defense&0.26*&0.599&\gr0.56&\gr0.554&\gr0.43&\gr0.509&0.04*&1.506\\
		&perspective-taking&\gr0.71&\gr0.248&\gr0.86&\gr0.235&0.40***&0.230&3.57***&0.278\\	
  
		\cline{1-2}
		\multirow{4}{1in}{Cognizance Guidance}
		&sympathy invoking&1.42*&0.153&1.49**&0.149&\gr1.27&\gr0.146&3.23***&0.185\\
		&evidence informing&2.02***&0.140&2.57***&0.138&2.45***&0.130&1.81**&0.189\\
		&gender educating&3.06***&0.190&3.02***&0.182&2.70***&0.195&4.98***&0.232\\
		&problem raising&2.11**&0.216&\gr1.19&\gr0.208&\gr1.24&\gr0.212&4.20***&0.241\\
		\hline
		\multicolumn{7}{l}{\textbf{Control Variables}}\\
            \hline
            \multirow{3}{1in}{Post Content} 
		&opinion (supporting women's needs)&0.57***&0.145&\gr0.77&\gr0.138&\gr0.90&\gr0.140&0.21***&0.170\\
		&incivility&1.54*&0.192&\gr1.04&\gr0.195&\gr0.81&\gr0.189&\gr1.26&\gr0.207\\
		&constructiveness&0.37***&0.162&0.53***&0.157&0.41***&0.155&\gr1.40&\gr0.210\\

		\cline{1-2}
            \multirow{2}{1in}{Post Characteristics}
		&post length&\gr1.00&\gr0.000&\gr1.00&\gr0.000&\gr1.00&\gr0.000&1.00*&0.001\\
		&hashtag\_num&\gr1.05&\gr0.082&1.18*&0.073&0.86*&0.069&1.33**&0.099\\

		\cline{1-2}
		\multirow{2}{1in}{Poster Info}
		&follower&1.00**&0.000&1.00***&0.000&1.00*&0.000&1.00***&0.000\\
		&following&1.00*&0.000&1.00**&0.000&1.00**&0.000&1.00**&0.000\\
		\hline\hline
           (Intercept)&&1.54*&0.190&\gr0.85&\gr0.181&3.78***&0.172&0.34***&0.230\\
           No. Observations&&\multicolumn{2}{c}{547}&\multicolumn{2}{c|}{547}&\multicolumn{2}{c}{547}&\multicolumn{2}{c}{547}\\
           Pseudo R-squ. &&\multicolumn{2}{c}{0.4730}&\multicolumn{2}{c|}{0.5968}&\multicolumn{2}{c}{0.4173}&\multicolumn{2}{c}{0.6704}\\
           Log Likelihood&&\multicolumn{2}{c}{-1009.0}&\multicolumn{2}{c|}{-1025.6}&\multicolumn{2}{c}{-1356.5}&\multicolumn{2}{c}{-789.75}\\
            \hline
	\end{tabular}}
 \label{TAB: response}
\end{table*}

\section{DISCUSSION}

This paper identifies a comprehensive taxonomy of discursive strategies in gender debate, and unveils their correlations with user participation and diverse dimensions of user response. This section contextualizes the findings within the existing literature, explores the opportunities and challenges of gender debate on social media, and puts forward design implications to enhance and facilitate gender-related discussions.
\subsection{From Everyday Feminism to Gender Debate}



\textit{Interweaving Conflicting Views and Constructive Insights.} Everyday online feminism has evolved into a significant form of advocating for women's daily needs \cite{yang2022research}. However, the enduring influence of patriarchal culture and the existence of knowledge gaps resulting from gender differences often lead to the neglect, opposition, and even stigmatization of online feminism within Chinese social media \cite{li2020collective,liu2016development}. Consequently, more intense narratives have emerged as a means of voicing these concerns, ultimately fueling the gender debate \cite{liyang}. To cultivate a more effective environment for everyday online feminism, it is essential to understand this context and provide appropriate support.

In terms of gender debate within everyday feminism, our findings demonstrated the prevalence and intertwined characteristics of uncivil and constructive discussions. We revealed a notable occurrence of uncivil (55.53\%) and constructive discourse (47.19\%), with a frequent coexistence of the two types of discourses in the same post (16.13\%) in Section \ref{Context Description}. On the one hand, incivility in feminist advocacy could emerge due to the strong emotions evoked by gender issues, leading individuals to express their frustrations, anger, or disappointment in an uncivil manner. Such language provides an avenue for emotional expression and facilitates connections among individuals who have shared experiences or perspectives \cite{schuster2017personal}. Additionally, incivility might serve as a form of resistance against oppressive narratives or systems. It could bring public attention to significant gender issues that might otherwise be overlooked \cite{cui2022comparative,mao2020feminist, d2013civil}. On the other hand, the presence of constructive information in everyday feminism assumes a vital function in advancing gender dialogue and effecting change. It acts as a medium for sharing knowledge, resources, and personal experiences, empowering individuals and fostering understanding between people with diverse perspectives \cite{oz2018twitter,papacharissi2004democracy,rowe2015civility}.

Further, we identified a range of uncivil and constructive strategies in Section \ref{RQ1}. For instance, \textit{sarcasm} and \textit{animal/sexist nomination} can be seen as a form of expression through which participants engage in confrontational discourse, potentially generating controversy and attracting attention to gender issues. We also witnessed the effective use of \textit{evidence informing} strategy to raise public awareness. The \textit{gender education} strategy played a pivotal role in bridging knowledge gaps, promoting empathy and enhancing understanding of women's challenges. These strategies enrich deliberative democracy \cite{bessette1980deliberative}, a classic political theory, by encouraging evidence-based discussions and equipping the public with knowledge. This expands the understanding and applicability of deliberative democracy in gender debate within everyday feminism.

The discussion surrounding the narrative of incivility and constructiveness necessitates a dialectical perspective that considers their limitations and complementarity. In the context of everyday feminism, certain strategies that frequently involve uncivil content, such as \textit{gender exclusion}, can be seen as a manifestation of radical feminism \cite{radical} to some extent. This strategy adopted radical means to exclude individuals of a specific gender from participating in issues related to another gender, thereby highlighting gender differences and power dynamics. Despite the potential risk of alienating others and cultivating a hostile atmosphere associated with this form of feminism \cite{firestone2015dialectic,echols1989daring}, it could also serve as a catalyst for raising awareness and emphasizing the need to address systemic issues \cite{willis1984radical}. Our findings provided empirical evidence that some strategies, like \textit{sympathy invoking} and \textit{suggestion}, increased incivility while also attracting constructive conversations. This finding supports the application of agonistic pluralism theory \cite{mouffe1999deliberative} to gender discourse, which views conflicts and opposition as essential for the healthy development of democracy. On the other hand, constructive information aligns to some extent with moderate feminism, emphasizing inclusivity and cooperation aimed at promoting more constructive and respectful conversations \cite{tzanakou2019moderate,lewis2019introduction}. Although moderate feminism may be more soothing, it might weaken the urgency and radical potential of feminist activism, potentially perpetuating the status quo \cite{budgeon2019resonance}. Therefore, incorporating a dialectical perspective allows us to acknowledge the strengths and weaknesses of each narrative while seeking a balanced and nuanced understanding.

Everyday feminism has become a significant global topic \cite{schuster2017personal,pruchniewska2019everyday}, encompassing various aspects of women's rights, such as education \cite{pedersen2021practical}, employment \cite{kelly2015feminist}, and safety (e.g., sexual harassment and violence) \cite{thrift2014yesallwomen}. Simultaneously, gender debate on social media has increasingly gained prominence \cite{peng2022digital,sagredos2022slut}. Understanding these debates might contribute to a deeper comprehension of gender conflicts, improve constructive communication regarding gender issues, and facilitate feminist advocacy on social media. Building upon this, our research revealed the interplay between constructive and uncivil dialogues, which holds the potential to serve as a meaningful reference and offer preliminary guidance for the extensive range of gender-related debates on social media. We call for future research to explore other context-specific discursive strategies and uncover their impacts on user participation and response.

\subsection{From Gender Debate to Discursive Strategies}
\subsubsection{Unveiling the Role of Discursive Strategies} 
\label{5.2.1}
Our findings in Section \ref{RQ2} indicated that \textit{role reversal} strategy was an infrequent but very powerful strategy, which significantly facilitated information dissemination. Berger and Milkman's research on the characteristics of viral content demonstrated that unexpectedly interesting and intensely emotional content tended to go viral \cite{berger2012makes}. \textit{Role reversal} might challenge traditional gender roles and provoke reflections on male privilege \cite{das2019gendered,metaxa2021image}, capturing users' interest, evoking strong emotions and increasing the likelihood of being shared. While leveraging \textit{role reversal} as a means to promote information sharing in gender debate could be effective, it is essential to handle uncivil content with caution to avoid controversy and negative reactions. 
 
Moreover, the counterproductive nature of strategies like offering \textit{suggestions} and engaging in \textit{perspective-taking}, which attract opposing views or foster uncivil content, is an intriguing finding. Upon reviewing posts with these strategies, we noticed that they sometimes intertwined with \textit{derogation} strategies (e.g., \textit{general name-calling} strategy), which might provoke backlash and hinder meaningful discourse. Additionally, \textit{suggestions} and \textit{perspective-taking} often represented only a few individuals' views, potentially causing dissatisfaction among entire population. Furthermore, some posts took on a didactic tone, appearing overly instructive like mansplaining \cite{koc2021online}, might not resonate well with intended audience. These factors, including mixed strategies, subjective content, and a prescriptive tone, might influence the effectiveness of posts in fostering meaningful and productive discourse on social media in China. On this note, we suggest social media platforms considering lightweight interventions before users posting, to enhance objectivity and comprehensiveness of posts. These interventions could include simple hints, such as demonstrating professional expertise or embracing dialectical thinking, though there might be a potential trade-off of reduced user engagement \cite{jahanbakhsh2021exploring}.




Another noteworthy finding revealed a positive association between \textit{overgeneralization} and \textit{recontextualization} strategies and user participation, yet their effect on user stance was controversial. According to social identity theory~\mbox{
\cite{tajfel1979integrative,tajfel2004social}}\hskip0pt
, individuals tend to associate themselves with specific groups and shape their self-identity through identification with those groups. Section \ref {4.3.3} demonstrated that \textit{recontextualization} strategy had the potential to draw in supporting views, probably appealing to in-group members who shared similar beliefs and perspectives in gender debate like advocating for women's rights. This might facilitate intergroup connections and promote a sense of identification
\cite{baughan2022shame,cosper2022patterns}, with the potential risk of limited viewpoint diversity and echo chamber effects \cite{cinelli2021echo}. Conversely, \textit{overgeneralization} strategy was associated with contrasting perspectives and probably engaged individuals with dissenting opinions, might appealing to out-group members. Although this strategy was possible to provide different viewpoints and challenges biases for other stakeholders in gender-related issues, it might also intensify conflicts such as incivility and misunderstandings \cite{ismail2020defying}. Further research is warranted to investigate gender differences in in-group and out-group gatherings associated with these strategies, including variations in gathering female and male users and group transformations during discussions.

\subsubsection{Comparing the Gendered Difference of Discursive Strategies} Section \ref{gender difference} showed that females exhibited higher levels of activity compared to males in gender debate of supporting women's needs. This might be attributed to the topic's direct relation to female rights or the larger female user base on Weibo \cite{weibouser}. We call for future cross-platform research to examine similar discussions on platforms with a higher male user presence. Furthermore, females commonly employed strategies like \textit{gender exclusion} and \textit{general name-calling}, highlighting gender differences and power dynamics, whereas males preferred informing strategies such as \textit{making suggestions} and \textit{identifying potential issues}. However, we suggest future work paying attention to not only the differences of discursive strategies between genders, but also the interaction and response of genders in these strategies, e.g., how one gender changes or amplify the views of another gender.


Section \ref{gender difference} revealed that men were found to be more inclined towards engaging in \textit{gender-related education} compared to women. However, upon further investigation, we discovered some posts were possibly written by females based on the context or previous posts of these accounts, despite the gender indicated in their profiles being male. This suggested our insights into gendered discursive strategies were tempered by potential fake gender profiles on social media, a limitation we have acknowledged in Section \ref{limitations}. In fact, we cannot determine whether this discrepancy was a deliberate act of lying about genders in the context of discussing gender issues like ``catfish'' \cite{magdy2017fake}, or simply a result of users inputting the opposite gender during profile creation. Therefore, future research could explore deeper into fake gender profiles such as users' motivations and their influence on public perceptions.

\subsection{Practical and Design Implications}
\subsubsection{Discovering Needs and Concerns: Implications for Social Support}
This work is grounded on social media, where users engage in self-initiated discussions, allowing the authentic grassroots voices of needs and concerns to emerge. Compared to traditional feminist methods, such as relying on institutional documents, academic musings, or interviews with a select group of veteran activists \cite{wu2019made}, spontaneous online expression through social media offers accessibility, community building, and information dissemination \cite{schuster2017personal}. For individual users, social media enables sharing their experiences and concerns, leading to emotional support, heightened awareness of women's needs, and collective empowerment. For policy makers, diverse perspectives in gender debate, as well as direct and real-time voices from the public on social media inform more relevant and impactful policy-making. Nevertheless, due to the content complexity generated by gender debate on social media, policymakers have to spend a significant amount of time organizing valuable opinions. To this end, it is warranted to integrate features on social media for conducting surveys, polls and feedback collection, and establishing dedicated channels for citizens to communicate directly with policy administration, making policy advocacy more responsive and inclusive. For companies, they could use social media to understand the public attitude on feminist issues, tailor their services or products accordingly (e.g., the demand for menstrual products on high-speed railway), and engage in corporate social responsibility initiatives. For researchers, the diverse discourse on social media provides invaluable data for studying contemporary feminist issues, trends, and public needs and challenges.


\subsubsection{Leveraging AI and User Feedback: Implications for Constructive Discussions}



The data pre-processing in Section \ref{processing} and findings in Section \ref{findings: strategies} implied that the current environment for discussions on gender issues was chaotic, due to many off-topic posts and lots of individuals posting to derogate probably for emotional expression rather than engaging in constructive dialogue. This excessively toxic narrative has a negative impact on public discussion, such as digital violence \cite{suarez2022toxic} and extreme feminism \cite{mo2022study}, which could create barriers to true equality between genders. This might also pose challenges for stake-holders such as policymakers and companies, who have to navigate through numerous posts to understand public needs and concerns. Thus, we envision the creation of an AI-supported infographic on the homepage of specialized topics dedicated to gender-related issues on social media. Specifically, considering chaotic content in gender debate, we suggest the infographic extracting and summarizing main arguments from posts related to gendered issues. These arguments could be organized in a hierarchical structure, with each argument accompanied by relevant examples. Additionally, gender-related discussions often involve contrasting viewpoints, such as those supporting or opposing women's needs. These viewpoints have the potential to provide a comprehensive understanding of public opinions and bridge the cognitive gap between genders. Therefore, when summarizing the main arguments in everyday feminism, these arguments could be categorized into ``arguments for women's needs'' and ``arguments for primary concerns.'' Moreover, it is valuable to update in real-time as the discussion evolves to ensure the information stays current. This tool might enable users to engage in more informed and productive discussions, and assist policymakers and companies in effectively identifying women's needs and concerns. However, though AI-supported infographic might provide valuable insights, further explorations are warranted to investigate potential limitations such as algorithmic bias.

Previous research has highlighted the interference and manipulation of users' debates on feminism by media companies and commercial interests \cite{mo2022study}. Media companies exploit ranking algorithms to deliberately amplify gender conflicts for their commercial objectives \cite{yang2022research,mcluhan2017medium}. For instance, \textit{recontextualization} strategy could draw user participation and uncivil content shown in Section \ref{RQ2}. If leveraged by media companies or commercial interests, these strategies are likely to exacerbate gender conflicts and limit effective communication. In this aspect, public wisdom might help alleviate this challenge. We suggest a user-driven labeling system on social media, which allows users to add tags to posts while browsing, such as whether they perceive the content as uncivil or constructive, supportive or unsupportive of women's needs, and recommended for females or males, to facilitate diverse information seeking. Based on user-driven labels, introducing corresponding sorting interfaces might help users filter and prioritize interested content, which potentially create a user-driven and gender-focused feed as a complement of the current sorting algorithms. In addition, user-generated tags provide one potential way for recommendation algorithms, through which users could obtain in-group and out-group information (e.g., recommendations related to supporting or opposing women's needs) to bridge gendered information gap and alleviate echo chamber. Nevertheless, it is warranted to investigate potential conflicts when recommending out-group opinions. There is also a risk of intentional collective behaviors \cite{suarez2022toxic}, particularly from certain groups that might deliberately tag uncivil or low-value content as intended for opposite gender.

\subsection{Limitations}
\label{limitations}
There are three limitations in this research: 1) Our analysis for discursive strategies is based on a sampled dataset of 1,928 entries, rather than the entirety of the data we collected. This sampling approach might introduce certain biases, potentially influencing the accuracy and comprehensiveness of our findings. 2) The data we accessed has undergone platform censorship, which means we might not be exposed to more uncivil content or some sensitive topics. This limitation could potentially hinder our understanding of the full spectrum of gender debate within everyday feminism on social media. 3) The reliability of our findings might be influenced by potential discrepancies in the gender information provided on social media profiles \cite{magdy2017fake}; 4) we do not investigate how different strategies in posts impact those in comments due to complex influencing factors (e.g., other comments within the same post). While these insights contribute to our understanding of the topic, they should be viewed as an initial exploration rather than definitive conclusions, due to the inherent limitations in accurately assessing gender representation in digital spaces.


\section{CONCLUSION}

There is a growing trend among users to utilize social media platforms as a means to advocate for women's daily needs. However, this has also given rise to debate surrounding gender issues. This work aims to explore user-developed discursive strategies in gender debate and their impact on user participation and user response within the realm of everyday feminism. To accomplish this, we collected a total of 38,636 posts and 187,539 comments from the Weibo platform and conducted a mixed-methods study. By employing an open coding approach, we developed a comprehensive taxonomy of user-generated strategies in gender debate, such as \textit{animal/sexist nomination} and \textit{gender education} strategies. Through regression analysis, we discovered these strategies' correlations with user participation and user response, such as the backfire effect of \textit{suggestion} strategy, and the low frequency but powerful \textit{role reversal} strategy. Our findings contribute to a deeper understanding of gender debate within everyday feminism and provide valuable insights for fostering constructive discussions in feminist advocacy on social media.

\bibliographystyle{ACM-Reference-Format}
\bibliography{sample-base}

\appendix
\section{Temporal Analysis of Gender Debate}\label{Temporal}

\begin{figure}
	\centering
		\includegraphics[scale=0.25]{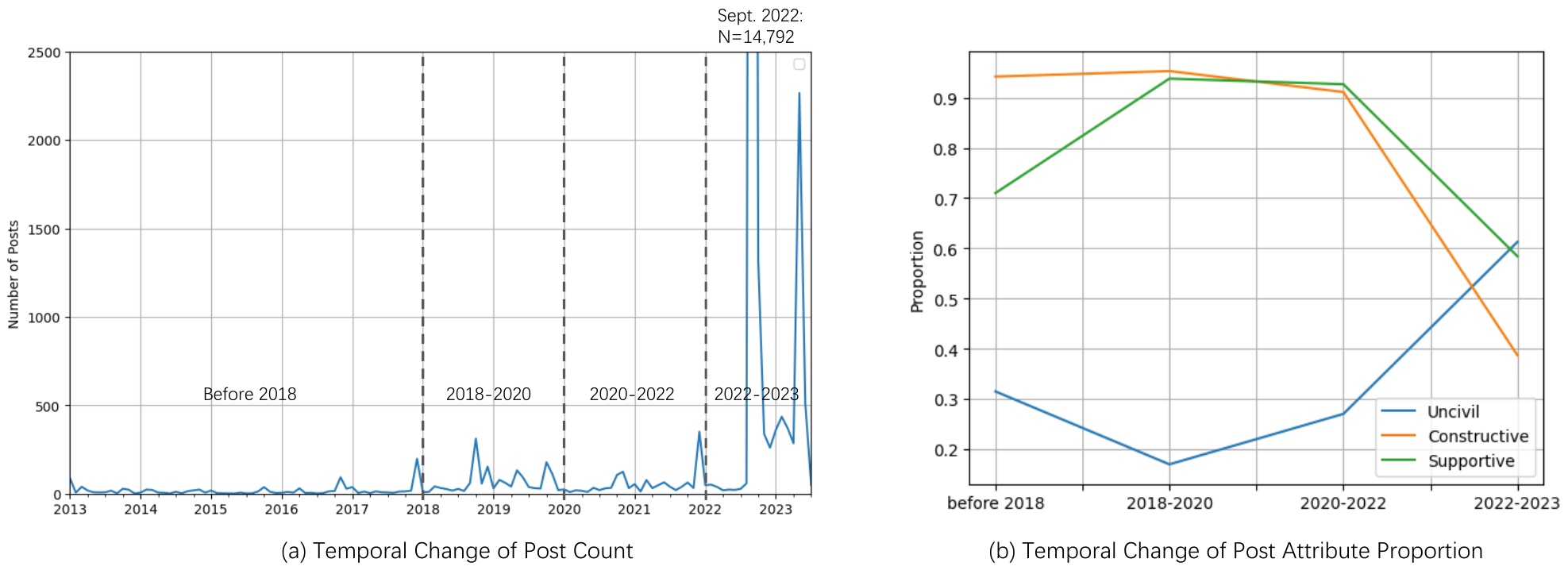}
	\caption{Temporal Analysis of Post Count and Attributes in Gender Debate}
        \Description{Figure 3 (a) shows the post volume was a relatively small number between 2013 and 2022, but there was an upsurge in September 2022 and the discussion became heated thereafter. Figure 3 (b) demonstrates that the proportions of uncivil posts greatly increased after 2018.}
	\label{FIG: temporal}
\end{figure}

Figure \ref{FIG: temporal} illustrated the temporal change of the post count and attributes in gender debate. Generally, the post volume in our dataset fluctuated within a relatively small number between 2013 to 2022 as demonstrated in Figure \ref{FIG: temporal} (a). A woman's complaint about the lack of menstrual products on high-speed railway in September 2022 triggered an upsurge in gender debate, and the discussion became heated thereafter. Based on the volume change, we divided the dataset into four temporal stages: \textit{before 2018} (N=1,000), \textit{2018-2020} (N=1,596), \textit{2020-2022} (N=1,284), and \textit{2022-2023} (N=21,268). Figure \ref{FIG: temporal} (b) described how the proportions of \textit{uncivil}, \textit{constructive} and \textit{supportive} posts changed over time. We observed that uncivil posts vastly increased in gender debate after 2018, indicating a potentially more polarized and toxic atmosphere in the discussion on everyday feminism~\cite{mo2022study,suarez2022toxic}.

\end{CJK}
\end{document}